\documentclass[10pt]{article}
\usepackage{graphicx}
\usepackage{amsmath}
\usepackage{amssymb}
\usepackage{caption2}
\begin{document}
\begin{center}
{\large {\bf \sc{  Analysis of the vector tetraquark states with  P-waves between the diquarks and antidiquarks via  the QCD sum rules }}} \\[2mm]
Zhi-Gang  Wang \footnote{E-mail: zgwang@aliyun.com.  }     \\
 Department of Physics, North China Electric Power University, Baoding 071003, P. R. China
\end{center}

\begin{abstract}
In this article, we introduce a P-wave between the diquark and antidiquark explicitly to construct the vector tetraquark currents,
 and study the  vector tetraquark states with the QCD sum rules systematically, and obtain the lowest vector tetraquark masses up to now.
  The present predictions support  assigning the
 $Y(4220/4260)$,  $Y(4320/4360)$, $Y(4390)$ and $Z(4250)$ to be the vector tetraquark   states with a relative P-wave between the diquark and antidiquark pair.
 \end{abstract}



\section{Introduction}
The attractive interactions induced by one-gluon exchange  favor  formation of
the diquarks in  color antitriplet, flavor antitriplet and spin singlet \cite{One-gluon}.
The diquarks $\varepsilon^{ijk}q^{T}_j C\Gamma q^{\prime}_k$ have  five  structures  in Dirac spinor space, where the $i$, $j$ and $k$ are color indexes,  $C\Gamma=C\gamma_5$, $C$, $C\gamma_\mu \gamma_5$,  $C\gamma_\mu $ and $C\sigma_{\mu\nu}$ for the scalar, pseudoscalar, vector, axialvector  and  tensor diquarks, respectively.  The favored or stable configurations are the scalar  and axialvector  diquark states from  the QCD sum rules \cite{WangDiquark,WangLDiquark,Tang-Diquark,Dosch-Diquark-1989}.
In the  non-relativistic quark model, an additional P-wave can change the parity by contributing a factor $(-)^L=-$,   where $L=1$ is the angular momentum.
The $C\gamma_5$ and $C\gamma_\mu$ diquark states have the spin-parity $J^P=0^+$ and $1^+$, respectively, while the $C$ and $C\gamma_\mu\gamma_5$ diquark states have the spin-parity $J^P=0^-$ and $1^-$, respectively. We can take the $C$ and $C\gamma_\mu\gamma_5$ diquark states as the P-wave excitations of the $C\gamma_5$(or $C\gamma_\alpha$) and $C\gamma_\mu$ diquark states, respectively, the net effects of the  P-waves   are embodied in the underlined  $\gamma_5$ in the  $C\gamma_5 \underline{\gamma_5} $ and $C\gamma_\mu \underline{\gamma_5} $ (or in the underlined  $\gamma^\alpha$ in the  $C\gamma_\alpha \underline{\gamma^\alpha} $).
We can also introduce the  P-wave explicitly in the  $C\gamma_5$ and $C\gamma_\mu$ diquark states and obtain the vector diquark states  $\varepsilon^{ijk}q^{T}_j C\gamma_5 \stackrel{\leftrightarrow}{\partial}_\mu q^{\prime}_k$ or the tensor diquark states  $\varepsilon^{ijk}q^{T}_j C\gamma_\mu \stackrel{\leftrightarrow}{\partial}_\nu q^{\prime}_k$, where the derivative  $\stackrel{\leftrightarrow}{\partial}_\mu=\stackrel{\rightarrow}{\partial}_\mu-\stackrel{\leftarrow}{\partial}_\mu$ embodies  the P-wave effects.
Thereafter, we will refer the $C\gamma_5$, $C\gamma_\mu $  diquark states as the S-wave diquark states and the $C$, $C\gamma_\mu\gamma_5$, $C\gamma_5\stackrel{\leftrightarrow}{\partial}_\mu$, $C\gamma_\mu\stackrel{\leftrightarrow}{\partial}_\nu$ diquark states as the P-wave diquark states.

 We can take the $C\gamma_5$ and $C\gamma_\mu$ diquark states and antidiquark states as the basic constituents to construct the scalar, axialvector and tensor tetraquark states,  for example, the $C\gamma_5 \otimes \gamma_5 C$, $C\gamma_\alpha \otimes \gamma^\alpha C$ type scalar tetraquark states \cite{WangMPLA}, the $C\gamma_\mu \otimes \gamma_5 C \pm C\gamma_5 \otimes \gamma_\mu C$ type axialvector tetraquark states \cite{Narison-X3872,WangHuangtao-2014-PRD}, the $C\gamma_\mu \otimes \gamma_\nu C+C\gamma_\nu \otimes \gamma_\mu C$ type tensor tetraquark states \cite{WangTetraquarkCTP}.

We can take  a S-wave and a P-wave diquark-antidiquark pair to construct the vector tetraquark states, or introduce an explicit P-wave in the S-wave diquark-antidiquark pair to construct the vector tetraquark states \cite{WangZG-CPL}. Experimentally, the $Y(4260)$ observed by the BaBar collaboration \cite{BaBar4260-0506}, the $Y(4220)$, $Y(4390)$ and $Y(4320)$ observed  by the BESIII collaboration  \cite{BES-Y4390,BES-Y4220-Y4320}, the $Y(4360)$, $Y(4660)$, $Y(4630)$ observed by the
  Belle collaboration \cite{Belle4660-0707,Belle4630-0807} are excellent candidates for the vector tetraquark states.  According to the analogous masses and widths,
   the $Y(4260)$ and $Y(4220)$ maybe the same particle, the $Y(4360)$ and $Y(4320)$ maybe the same particle, the $Y(4660)$ and $Y(4630)$ maybe the same particle.
  In Table 1, we present the possible assignments of the $Y$ states as vector tetraquark states based on the QCD sum rules \cite{Nielsen-4260-4460,ChenZhu,WangY4360Y4660-1803,Wang-tetra-formula,WangEPJC-1601,ZhangHuang-PRD,ZhangHuang-JHEP,WZG1809}.
  In Refs.\cite{Nielsen-4260-4460,ChenZhu,WangY4360Y4660-1803}, the same interpolating currents lead to quite different assignments, because different input parameters are chosen  at the QCD side of the QCD sum rules. In the QCD sum rules for the hidden-charm (or hidden-bottom) tetraquark states and molecular states, the integrals
 \begin{eqnarray}
 \int_{4m_Q^2(\mu)}^{s_0} ds \rho_{QCD}(s,\mu)\exp\left(-\frac{s}{T^2} \right)\, ,
 \end{eqnarray}
are sensitive to the heavy quark masses $m_Q$, where the $\rho_{QCD}(s,\mu)$ are  the QCD spectral densities, the $T^2$ are the Borel parameters, the $s_0$ are the continuum thresholds parameters.
Variations of the heavy quark masses or the energy scales $\mu$ lead to variations of integral ranges $4m_Q^2-s_0$ of the variable  $ds$ besides the QCD spectral densities $\rho_{QCD}(s,\mu)$, therefore variations of the Borel windows and predicted masses and pole residues. In Refs.\cite{Wang-tetra-formula,WangHuang-molecule}, we suggest an energy scale formula $\mu=\sqrt{M^2_{X/Y/Z}-(2{\mathbb{M}}_Q)^2}$ with the effective $Q$-quark masses  ${\mathbb{M}}_Q$ to determine the ideal energy scales of the QCD spectral densities.   Compared to the old predictions in Ref.\cite{Wang-tetra-formula}, the new predictions based on detailed analysis  with the   updated   parameters  are preferred \cite{WangY4360Y4660-1803}. The $C\gamma_5\otimes\partial_\mu\otimes\gamma_5C$ type interpolating currents chosen in Refs.\cite{ZhangHuang-PRD,ZhangHuang-JHEP} have no definite charge conjugation.  In Ref.\cite{WZG1809}, we   construct  the $C\gamma_5\otimes\stackrel{\leftrightarrow}{\partial}_\mu\otimes \gamma_5C$ type interpolating  current with $J^{PC}=1^{--}$ to study the lowest vector tetraquark state with the QCD sum rules  by carrying out the operator product expansion up to the vacuum condensates of dimension 10, and use the modified  energy scale formula $\mu=\sqrt{M^2_{X/Y/Z}-(2{\mathbb{M}}_c+0.5\,\rm{GeV})^2}=\sqrt{M^2_{X/Y/Z}-(4.1\,\rm{GeV})^2}$  to determine the ideal  energy scale of the QCD spectral density, where we have assumed  that an additional P-wave costs about $0.5\,\rm{GeV}$.

  In the four-quark system $qQ\bar{q}^{\prime}\bar{Q}$,
 the $Q$-quark serves as a static well potential and  attracts  the light quark $q$  to form a heavy diquark in  color antitriplet,
while the $\bar{Q}$-quark serves  as another static well potential and attracts the light antiquark $\bar{q}^\prime$  to form a heavy antidiquark  in  color triplet \cite{Wang-tetra-formula,Wang-tetra-NPA}.
 The   diquark and antidiquark  attract  each other    to form a compact tetraquark state \cite{Wang-tetra-formula,Wang-tetra-NPA},
the two heavy quarks $Q$ and $\bar{Q}$ stabilize the tetraquark state, just like the $\mu^+$ and $\mu^-$ stabilize  the $\mu^+e^-\mu^-e^+$ system  \cite{Brodsky-PRL}. The tetraquark states    are characterized by the effective heavy quark masses ${\mathbb{M}}_Q$ and the virtuality $V=\sqrt{M^2_{X/Y/Z}-(2{\mathbb{M}}_Q)^2}$. If there is an additional P-wave between the diquark and antidiquark, in other words, between the heavy quark $Q$ and heavy antiquark  $\bar{Q}$, the virtuality should be modified to be $V=\sqrt{M^2_{X/Y/Z}-(2{\mathbb{M}}_Q+0.5\,\rm{GeV})^2}$, therefore the energy scale formula $\mu=V$ is also modified.
  For the $C\gamma_5 \otimes \gamma_5\gamma_\mu C$-type and $C \otimes \gamma_\mu C$-type vector tetraquark states, the relative P-waves lie in the P-wave diquarks or antidiquarks, not lie between the diquark and antidiquark, the energy scale formula $\mu=\sqrt{M^2_{X/Y/Z}-(2{\mathbb{M}}_Q)^2}$ works \cite{WangY4360Y4660-1803,Wang-tetra-formula,WangEPJC-1601}.

\begin{table}
\begin{center}
\begin{tabular}{|c|c|c|c|c|c|c|c|}\hline\hline
           &Structures                                    &Constituents                           &OPE\,(No)    &mass(GeV)      &References   \\ \hline
$Y(4660)$  &$C\gamma_5 \otimes \gamma_5\gamma_\mu C$      &$c\bar{c}s\bar{s}$                     &$8\,(7)$     &$4.65$         &\cite{Nielsen-4260-4460}  \\ \hline
$Y(4660)$  &$C\gamma_5 \otimes \gamma_5\gamma_\mu C$      &$c\bar{c}q\bar{q}$                     &$8\,(7)$     &$4.64$         &\cite{ChenZhu}  \\ \hline
$Y(4360)$  &$C\gamma_5 \otimes \gamma_5\gamma_\mu C$      &$c\bar{c}q\bar{q}$                     &$10$         &$4.34$         &\cite{WangY4360Y4660-1803}  \\ \hline

$Y(4660)$  &$C \otimes \gamma_\mu C$                      &$c\bar{c}s\bar{s}/c\bar{c}q\bar{q}$    &$10$         &$4.66/4.59$    &\cite{WangY4360Y4660-1803}  \\ \hline
$Y(4660)$  &$C \otimes \gamma_\mu C$                      &$c\bar{c}s\bar{s}/c\bar{c}q\bar{q}$    &$10$         &$4.70/4.66$    &\cite{Wang-tetra-formula}  \\ \hline

$Y(4660)$  &$C\gamma_\mu \otimes \gamma_\nu C-C\gamma_\nu \otimes \gamma_\mu C$   &$c\bar{c}q\bar{q}$  &$10$    &$4.66$         &\cite{WangEPJC-1601}  \\ \hline

$Y(4660)$  &$C\gamma_5\otimes\partial_\mu\otimes\gamma_5C$&$c\bar{c}s\bar{s}$                     &$6$          &$4.69$         &\cite{ZhangHuang-PRD}  \\ \hline

$Y(4360)$  &$C\gamma_5\otimes\partial_\mu\otimes\gamma_5C$&$c\bar{c}q\bar{q}$                     &$6$          &$4.32$         &\cite{ZhangHuang-JHEP}  \\ \hline

$Y(4260)$  &$C\gamma_5\otimes\stackrel{\leftrightarrow}{\partial}_\mu\otimes\gamma_5C$  &$c\bar{c}q\bar{q}$     &$10$      &$4.24$    &\cite{WZG1809}  \\ \hline \hline
\end{tabular}
\end{center}
\caption{ The OPE denotes  truncations of  the operator product expansion up to the vacuum condensates of dimension $n$, the No denotes the vacuum condensates of dimension $n^\prime$ are not included.   }
\end{table}

In this article, we extend our previous  work \cite{WZG1809} to study other vector tetraquark states with an explicit relative P-wave between the diquark and antidiquark with the QCD sum rules in a systematic way.
In the type-II diquark-antidiquark model \cite{Maiani-II-type},  L. Maiani et al assign  the $Y(4008)$, $Y(4260)$, $Y(4290/4220)$  and $Y(4630)$ to be  the four ground  states with  $L=1$ based on the effective  Hamiltonian with the spin-spin and spin-orbit  interactions by neglecting  the spin-spin interactions between the quarks and antiquarks. In Ref.\cite{Ali-Maiani-Y}, A. Ali et al incorporate the dominant spin-spin, spin-orbit and tensor interactions, and observe that the preferred  assignments of the ground state tetraquark states with  $L=1$ are the $Y(4220)$, $Y(4330)$, $Y(4390)$, $Y(4660)$.
In the diquark-antidiquark model, the quantum numbers of the $Y$ states are shown explicitly in Table 2, where the $L$ is the angular momentum between the diquark and antidiquark, $\vec{S}=\vec{S}_{qc}+ \vec{S}_{\bar{q}\bar{c}}$, $\vec{J}=\vec{S}+ \vec{L}$.
In this article, we reexamine those assignments based on the QCD sum rules, which is a powerful theoretical tool in studying the exotic $X$, $Y$, $Z$ particles.  In the QCD sum rules, the input parameters are the vacuum condensates and quark masses, which have universal values.

In the isospin limit, the vector tetraquark states with the symbolic quark constituents
\begin{eqnarray}
I=1&:& c\bar{c} u\bar{d}\, , \, \, \,  c\bar{c} \frac{u\bar{u}-d\bar{d}}{\sqrt{2}}\, , \, \, \, c\bar{c} d\bar{u}\, , \nonumber\\
I=0&:& c\bar{c} \frac{u\bar{u}+d\bar{d}}{\sqrt{2}}\, ,
\end{eqnarray}
have degenerate masses. In this article, we study the $c\bar{c} u\bar{d}$ tetraquark states for simplicity.
Now we construct the interpolating currents according the quantum numbers shown in Table 2,
\begin{eqnarray}
J^1_\mu(x)&=&\frac{\varepsilon^{ijk}\varepsilon^{imn}}{\sqrt{2}}u^{Tj}(x)C\gamma_5 c^k(x)\stackrel{\leftrightarrow}{\partial}_\mu \bar{d}^m(x)\gamma_5 C \bar{c}^{Tn}(x) \, ,  \\
J^2_\mu(x)&=&\frac{\varepsilon^{ijk}\varepsilon^{imn}}{\sqrt{2}}u^{Tj}(x)C\gamma_\alpha c^k(x)\stackrel{\leftrightarrow}{\partial}_\mu \bar{d}^m(x)\gamma^\alpha C \bar{c}^{Tn}(x) \, ,  \\
J^3_\mu(x)&=&\frac{\varepsilon^{ijk}\varepsilon^{imn}}{2}\left[u^{Tj}(x)C\gamma_\mu c^k(x)\stackrel{\leftrightarrow}{\partial}_\alpha \bar{d}^m(x)\gamma^\alpha C \bar{c}^{Tn}(x) \right.\nonumber\\
&&\left.+u^{Tj}(x)C\gamma^\alpha c^k(x)\stackrel{\leftrightarrow}{\partial}_\alpha \bar{d}^m(x)\gamma_\mu C \bar{c}^{Tn}(x)\right]\, ,  \\
J_{\mu\nu}(x)&=&\frac{\varepsilon^{ijk}\varepsilon^{imn}}{2\sqrt{2}}\left[u^{Tj}(x)C\gamma_5 c^k(x)\stackrel{\leftrightarrow}{\partial}_\mu \bar{d}^m(x)\gamma_\nu C \bar{c}^{Tn}(x) \right.\nonumber\\
&&+u^{Tj}(x)C\gamma_\nu c^k(x)\stackrel{\leftrightarrow}{\partial}_\mu \bar{d}^m(x)\gamma_5 C \bar{c}^{Tn}(x) \nonumber  \\
&&-u^{Tj}(x)C\gamma_5 c^k(x)\stackrel{\leftrightarrow}{\partial}_\nu \bar{d}^m(x)\gamma_\mu C \bar{c}^{Tn}(x) \nonumber \\
&&\left.-u^{Tj}(x)C\gamma_\mu c^k(x)\stackrel{\leftrightarrow}{\partial}_\nu \bar{d}^m(x)\gamma_5 C \bar{c}^{Tn}(x)\right]\, .
\end{eqnarray}

\begin{table}
\begin{center}
\begin{tabular}{|c|c|c|c|c|c|c|c|}\hline\hline
$|S_{qc}, S_{\bar{q}\bar{c}}; S, L; J\rangle$                                   &\cite{Maiani-II-type} &\cite{Ali-Maiani-Y} &Currents \\ \hline

$|0, 0; 0, 1; 1\rangle$                                                         &$Y(4008)$             &$Y(4220)$           &$J_\mu^1(x)$   \\ \hline

$\frac{1}{\sqrt{2}}\left(|1, 0; 1, 1; 1\rangle+|0, 1; 1, 1; 1\rangle\right)$    &$Y(4260)$             &$Y(4330)$           &$J_{\mu\nu}(x)$   \\ \hline

$|1, 1; 0, 1; 1\rangle$                                                         &$Y(4290/4220)$        &$Y(4390)$           &$J_\mu^2(x)$  \\ \hline

$|1, 1; 2, 1; 1\rangle$                                                         &$Y(4630)$             &$Y(4660)$           &$J_\mu^3(x)$  \\ \hline

$|1, 1; 2, 3; 1\rangle$                                                         &                      &                    &\\ \hline  \hline
\end{tabular}
\end{center}
\caption{ The vector tetraquark states, possible assignments  and the corresponding vector tetraquark currents, where the mixing effects are neglected.   }
\end{table}

In this article, we choose  the currents $J^1_\mu(x)$, $J^2_\mu(x)$, $J^3_\mu(x)$ and $J_{\mu\nu}(x)$ to study the vector tetraquark states with the QCD sum rules systematically  by calculating  the vacuum condensates up to dimension 10 in a consistent way in the operator product expansion, and use the modified  energy scale formula $\mu=\sqrt{M^2_{X/Y/Z}-(2{\mathbb{M}}_c+0.5\,\rm{GeV})^2}=\sqrt{M^2_{X/Y/Z}-(4.1\,\rm{GeV})^2}$  to determine the ideal  energy scales of the QCD spectral densities, and reexamine the possible assignments of the $Y$ states.

The article is arranged as follows:  we derive the QCD sum rules for the masses and pole residues  of  the vector   tetraquark states in section 2; in section 3, we   present the numerical results and discussions; section 4 is reserved for our conclusion.

\section{QCD sum rules for  the  vector tetraquark states}
In the following, we write down  the two-point correlation functions  $\Pi_{\mu\nu}(p)$ and $\Pi_{\mu\nu\alpha\beta}(p)$ in the QCD sum rules,
\begin{eqnarray}
\Pi_{\mu\nu}(p)&=&i\int d^4x e^{ip \cdot x} \langle0|T\left\{J_\mu(x)J_\nu^{\dagger}(0)\right\}|0\rangle \, , \\
\Pi_{\mu\nu\alpha\beta}(p)&=&i\int d^4x e^{ip \cdot x} \langle0|T\left\{J_{\mu\nu}(x)J_{\alpha\beta}^{\dagger}(0)\right\}|0\rangle \, ,
\end{eqnarray}
where $J_\mu(x)=J_\mu^1(x)$, $J_\mu^2(x)$ and $J_\mu^3(x)$.
Under charge conjugation transform $\widehat{C}$, the currents $J_\mu(x)$ and $J_{\mu\nu}(x)$ have the property,
\begin{eqnarray}
\widehat{C}J_{\mu}(x)\widehat{C}^{-1}&=&- J_{\mu}(x) \, , \nonumber\\
\widehat{C}J_{\mu\nu}(x)\widehat{C}^{-1}&=&- J_{\mu\nu}(x) \, ,
\end{eqnarray}
the currents have definite charge conjugation.

At the phenomenological side, we can insert  a complete set of intermediate hadronic states with
the same quantum numbers as the current operators $J_\mu(x)$ and $J_{\mu\nu}(x)$ into the
correlation functions  $\Pi_{\mu\nu}(p)$ and $\Pi_{\mu\nu\alpha\beta}(p)$ respectively   to obtain the hadronic representation
\cite{SVZ79,Reinders85}. After isolating the ground state  vector tetraquark  contributions,  we obtain the results,
\begin{eqnarray}
\Pi_{\mu\nu}(p)&=&\frac{\lambda_{Y}^2}{M_{Y}^2-p^2}\left(-g_{\mu\nu} +\frac{p_\mu p_\nu}{p^2}\right) + \cdots \, \, ,\nonumber\\
&=&\Pi_Y(p^2)\left(-g_{\mu\nu} +\frac{p_\mu p_\nu}{p^2}\right) +\cdots  \, , \\
\Pi_{\mu\nu\alpha\beta}(p)&=&\frac{\lambda_{ Y}^2}{M_{Y}^2\left(M_{Y}^2-p^2\right)}\left(p^2g_{\mu\alpha}g_{\nu\beta} -p^2g_{\mu\beta}g_{\nu\alpha} -g_{\mu\alpha}p_{\nu}p_{\beta}-g_{\nu\beta}p_{\mu}p_{\alpha}+g_{\mu\beta}p_{\nu}p_{\alpha}+g_{\nu\alpha}p_{\mu}p_{\beta}\right) \nonumber\\
&&+\frac{\lambda_{ Z}^2}{M_{Z}^2\left(M_{Z}^2-p^2\right)}\left( -g_{\mu\alpha}p_{\nu}p_{\beta}-g_{\nu\beta}p_{\mu}p_{\alpha}+g_{\mu\beta}p_{\nu}p_{\alpha}+g_{\nu\alpha}p_{\mu}p_{\beta}\right) +\cdots \, , \nonumber\\
&=&\widetilde{\Pi}_{Y}(p^2)\left(p^2g_{\mu\alpha}g_{\nu\beta} -p^2g_{\mu\beta}g_{\nu\alpha} -g_{\mu\alpha}p_{\nu}p_{\beta}-g_{\nu\beta}p_{\mu}p_{\alpha}+g_{\mu\beta}p_{\nu}p_{\alpha}+g_{\nu\alpha}p_{\mu}p_{\beta}\right) \nonumber\\
&&+\widetilde{\Pi}_{Z}(p^2)\left( -g_{\mu\alpha}p_{\nu}p_{\beta}-g_{\nu\beta}p_{\mu}p_{\alpha}+g_{\mu\beta}p_{\nu}p_{\alpha}+g_{\nu\alpha}p_{\mu}p_{\beta}\right) \, .
\end{eqnarray}
where the pole residues  $\lambda_{Y}$ and $\lambda_{Z}$ are  defined by
\begin{eqnarray}
\langle 0|J_\mu(0)|Y(p)\rangle &=&\lambda_{Y} \,\varepsilon_\mu \, , \nonumber\\
  \langle 0|J_{\mu\nu}(0)|Y(p)\rangle &=& \frac{\lambda_{Y}}{M_{Y}} \, \varepsilon_{\mu\nu\alpha\beta} \, \varepsilon^{\alpha}p^{\beta}\, , \nonumber\\
 \langle 0|J_{\mu\nu}(0)|Z(p)\rangle &=& \frac{\lambda_{Z}}{M_{Z}} \left(\varepsilon_{\mu}p_{\nu}-\varepsilon_{\nu}p_{\mu} \right)\, ,
\end{eqnarray}
the $\varepsilon_\mu$ are the polarization vectors  of the  vector tetraquark states $Y$ and axialvector tetraquark states $Z$ with the $J^{PC}=1^{--}$ and $1^{+-}$, respectively.
Now we project out the components $\Pi_{Y}(p^2)$ and $\Pi_{Z}(p^2)$ by introducing the operators $P_{Y}^{\mu\nu\alpha\beta}$ and $P_{Z}^{\mu\nu\alpha\beta}$,
\begin{eqnarray}
\Pi_{Y}(p^2)&=&p^2\widetilde{\Pi}_{Y}(p^2)=P_{Y}^{\mu\nu\alpha\beta}\Pi_{\mu\nu\alpha\beta}(p) \, , \nonumber\\
\Pi_{Z}(p^2)&=&p^2\widetilde{\Pi}_{Z}(p^2)=P_{Z}^{\mu\nu\alpha\beta}\Pi_{\mu\nu\alpha\beta}(p) \, ,
\end{eqnarray}
where
\begin{eqnarray}
P_{Y}^{\mu\nu\alpha\beta}&=&\frac{1}{6}\left( g^{\mu\alpha}-\frac{p^\mu p^\alpha}{p^2}\right)\left( g^{\nu\beta}-\frac{p^\nu p^\beta}{p^2}\right)\, , \nonumber\\
P_{Z}^{\mu\nu\alpha\beta}&=&\frac{1}{6}\left( g^{\mu\alpha}-\frac{p^\mu p^\alpha}{p^2}\right)\left( g^{\nu\beta}-\frac{p^\nu p^\beta}{p^2}\right)-\frac{1}{6}g^{\mu\alpha}g^{\nu\beta}\, .
\end{eqnarray}
 In this article, we choose the components $\Pi_{Y}(p^2)$ to study the vector tetraquark states.

 At the QCD side, we carry out the operator product expansion up to the vacuum condensates of   dimension-10, and take into account the vacuum condensates which are
vacuum expectations  of the operators  of the orders $\mathcal{O}( \alpha_s^{k})$ with $k\leq 1$ consistently. For the technical details, one can consult Refs.\cite{WangHuangtao-2014-PRD,WZG1809}.
 Once analytical expressions of the QCD spectral densities  are obtained,  we can take the
quark-hadron duality below the continuum threshold  $s_0$ and perform Borel transform  with respect to
the variable $P^2=-p^2$ to obtain  the  QCD sum rules:
\begin{eqnarray}
\lambda^2_{Y}\, \exp\left(-\frac{M^2_{Y}}{T^2}\right)= \int_{4m_c^2}^{s_0} ds\, \rho(s) \, \exp\left(-\frac{s}{T^2}\right) \, ,
\end{eqnarray}
where
\begin{eqnarray}
\rho(s)&=&\rho_{0}(s)+\rho_{3}(s) +\rho_{4}(s)+\rho_{5}(s)+\rho_{6}(s)+\rho_{7}(s) +\rho_{8}(s)+\rho_{10}(s)\, ,
\end{eqnarray}
 the subscripts $i$ in the spectral densities $\rho_{i}(s)$ denote the dimensions of the vacuum condensates,
\begin{eqnarray}
\rho_{3}(s)&\propto& \langle\bar{q}q\rangle\, ,\nonumber\\
\rho_{4}(s)&\propto& \langle\frac{\alpha_{s}GG}{\pi}\rangle\, ,\nonumber\\
\rho_{5}(s)&\propto& \langle\bar{q}g_s\sigma Gq\rangle\, ,\nonumber\\
\rho_{6}(s)&\propto& \langle\bar{q}q\rangle^2\, ,\nonumber\\
\rho_{7}(s)&\propto& \langle\bar{q}q\rangle\langle\frac{\alpha_{s}GG}{\pi}\rangle\, ,\nonumber\\
\rho_{8}(s)&\propto& \langle\bar{q}q\rangle\langle\bar{q}g_s\sigma Gq\rangle\, ,\nonumber\\
\rho_{10}(s)&\propto&  \langle\bar{q}g_s\sigma Gq\rangle^2\, ,\, \langle\bar{q}q\rangle^2\langle\frac{\alpha_{s}GG}{\pi}\rangle\, ,
\end{eqnarray}
the lengthy expressions of the QCD spectral densities are neglected for simplicity, the interested readers can obtain them through my E-mail.
The relatively simple expressions of the QCD spectral densities for the current $J^1_\mu(x)$ are presented  in Ref.\cite{WZG1809}.  For the currents $J^1_\mu(x)$ and $J^2_\mu(x)$, we take into account
all the contributions $\rho_{i}(s)$ with $i=0$, $3$, $4$, $5$, $6$, $7$, $8$, $10$.  In calculations, we observe that the contributions of the vacuum condensates $\langle\frac{\alpha_{s}GG}{\pi}\rangle$, $\langle\bar{q}q\rangle\langle\frac{\alpha_{s}GG}{\pi}\rangle$ and $\langle\bar{q}q\rangle^2\langle\frac{\alpha_{s}GG}{\pi}\rangle$ play a minor important role in the Borel windows, the predicted masses are almost the same if we neglect their contributions, furthermore, they also play a minor  important role in  determining the Borel windows. So we neglect the contributions of the vacuum condensates $\langle\frac{\alpha_{s}GG}{\pi}\rangle$, $\langle\bar{q}q\rangle\langle\frac{\alpha_{s}GG}{\pi}\rangle$ and $\langle\bar{q}q\rangle^2\langle\frac{\alpha_{s}GG}{\pi}\rangle$ in the QCD spectral densities for the currents $J^3_\mu(x)$ and $J_{\mu\nu}(x)$
due to the formidable  calculations in the operator product expansion.

For the current $J_\mu(x)=J^1_\mu(x)$, the correlation function $\Pi_{\mu\nu}(p)$ can be written as
\begin{eqnarray}
\Pi_{\mu\nu}(p)&=&-\frac{i\varepsilon^{ijk}\varepsilon^{imn}\varepsilon^{i^{\prime}j^{\prime}k^{\prime}}\varepsilon^{i^{\prime}m^{\prime}n^{\prime}}}{2}\int d^4x e^{ip \cdot x}   \nonumber\\
&&\left\{{\rm Tr}\left[\gamma_5 C^{kk^{\prime}}(x)\gamma_5 CS^{jj^{\prime}T}(x)C\right] \partial_\mu \partial_\nu{\rm Tr}\left[ \gamma_5 C^{n^{\prime}n}(-x)\gamma_5 C S^{m^{\prime}mT}(-x)C\right] \right. \nonumber\\
&&-\partial_\mu{\rm Tr}\left[ \gamma_5 C^{kk^{\prime}}(x)\gamma_5 CS^{jj^{\prime}T}(x)C\right] \partial_\nu{\rm Tr}\left[ \gamma_5 C^{n^{\prime}n}(-x)\gamma_5 C S^{m^{\prime}mT}(-x)C\right] \nonumber\\
&&-\partial_\nu{\rm Tr}\left[ \gamma_5 C^{kk^{\prime}}(x)\gamma_5 CS^{jj^{\prime}T}(x)C\right] \partial_\mu{\rm Tr}\left[ \gamma_5 C^{n^{\prime}n}(-x)\gamma_5 C S^{m^{\prime}mT}(-x)C\right] \nonumber\\
 &&\left.+\partial_\mu \partial_\nu{\rm Tr}\left[ \gamma_5 C^{kk^{\prime}}(x)\gamma_5 CS^{jj^{\prime}T}(x)C\right] {\rm Tr}\left[  \gamma_5C^{n^{\prime}n}(-x)\gamma_5 C S^{m^{\prime}mT}(-x)C\right] \right\}\, ,
\end{eqnarray}
where  the $S_{ij}(x)$ and $C_{ij}(x)$ are the full $u/d$ and $c$ quark propagators, respectively. In other words, $\Pi_{\mu\nu}(p)\propto \int d^4x\, e^{ip \cdot x} \,\rm{Tr}[\cdots]\times \rm{Tr}[\cdots]$. The first $\rm{Tr}[\cdots]$ contains quark lines for the diquark state, while the second  $\rm{Tr}[\cdots]$ contains quark lines for the antidiquark state. The contributions originate from the interactions between the quark lines in the first $\rm{Tr}[\cdots]$ or in the second $\rm{Tr}[\cdots]$ are factorizible,
while the  contributions originate from the interactions between the quark lines in the first $\rm{Tr}[\cdots]$ and in the second $\rm{Tr}[\cdots]$ are non-factorizible. In other words, the inner-diquark interactions are factorizible, while the inter-diquark interactions are non-factorizible. In Refs.\cite{Maiani-1712,Esposito-1807}, the authors  assume that there exists a repulsive barrier with finite width between the  diquarks and antidiquarks in the tetraquark states, which
 can answer satisfactorily some long standing questions challenging the diquark-antidiquark model of exotic resonances, for example, the non-observation of charged
partners $X^\pm$ of the $X(3872)$ and the absence of a hyperfine splitting between two different neutral
states, the tetraquark states decay more copiously into
open flavor mesons rather than quarkonia. In the present work, we observe that the dominant contributions come from the factorizible interactions (or Feynman-diagrams), the non-factorizible interactions (or Feynman-diagrams) play a much less  important role, which are consistent with the inter-diquark  barrier introduced in Refs.\cite{Maiani-1712,Esposito-1807}.   The finite potential barrier between diquarks could make the tetraquark state metastable against collapse and fall
apart decay, which happens if one of the quarks tunnels towards the other side. The non-factorizible interactions correspond to the tunneling effects in Refs.\cite{Maiani-1712,Esposito-1807} qualitatively. The conclusion survives for other currents.

We derive Eq.(15) with respect to  $\tau=\frac{1}{T^2}$, then eliminate the
 pole residues  $\lambda_{Y}$, and obtain the QCD sum rules for
 the masses of the vector   tetraquark states,
 \begin{eqnarray}
 M^2_{Y}&=& -\frac{\int_{4m_c^2}^{s_0} ds\frac{d}{d \tau}\rho(s)\exp\left(-\tau s \right)}{\int_{4m_c^2}^{s_0} ds \rho(s)\exp\left(-\tau s\right)}\, .
\end{eqnarray}

\section{Numerical results and discussions}
We take  the standard values of the vacuum condensates $\langle
\bar{q}q \rangle=-(0.24\pm 0.01\, \rm{GeV})^3$,   $\langle
\bar{q}g_s\sigma G q \rangle=m_0^2\langle \bar{q}q \rangle$,
$m_0^2=(0.8 \pm 0.1)\,\rm{GeV}^2$,  $\langle \frac{\alpha_s
GG}{\pi}\rangle=(0.33\,\rm{GeV})^4 $    at the energy scale  $\mu=1\, \rm{GeV}$
\cite{SVZ79,Reinders85,Colangelo-Review}, and choose the $\overline{MS}$ mass $m_{c}(m_c)=(1.275\pm0.025)\,\rm{GeV}$ from the Particle Data Group \cite{PDG}, and set $m_u=m_d=0$.
Moreover, we take into account the energy-scale dependence of  the input parameters on the QCD side,
\begin{eqnarray}
\langle\bar{q}q \rangle(\mu)&=&\langle\bar{q}q \rangle(Q)\left[\frac{\alpha_{s}(Q)}{\alpha_{s}(\mu)}\right]^{\frac{12}{25}}\, , \nonumber\\
 \langle\bar{q}g_s \sigma Gq \rangle(\mu)&=&\langle\bar{q}g_s \sigma Gq \rangle(Q)\left[\frac{\alpha_{s}(Q)}{\alpha_{s}(\mu)}\right]^{\frac{2}{25}}\, , \nonumber\\ m_c(\mu)&=&m_c(m_c)\left[\frac{\alpha_{s}(\mu)}{\alpha_{s}(m_c)}\right]^{\frac{12}{25}} \, ,\nonumber\\
\alpha_s(\mu)&=&\frac{1}{b_0t}\left[1-\frac{b_1}{b_0^2}\frac{\log t}{t} +\frac{b_1^2(\log^2{t}-\log{t}-1)+b_0b_2}{b_0^4t^2}\right]\, ,
\end{eqnarray}
   where $t=\log \frac{\mu^2}{\Lambda^2}$, $b_0=\frac{33-2n_f}{12\pi}$, $b_1=\frac{153-19n_f}{24\pi^2}$, $b_2=\frac{2857-\frac{5033}{9}n_f+\frac{325}{27}n_f^2}{128\pi^3}$,  $\Lambda=210\,\rm{MeV}$, $292\,\rm{MeV}$  and  $332\,\rm{MeV}$ for the flavors  $n_f=5$, $4$ and $3$, respectively \cite{PDG,Narison-mix}, and evolve all the input parameters to the ideal  energy scales  $\mu$ to extract the masses  of the
   vector tetraquark states, in other works,  choose the ideal  energy scales   $\mu$ to satisfy the relation $M^2_{X/Y/Z}=\mu^2+(4.1\,\rm{GeV})^2$ \cite{WZG1809}.

In Ref.\cite{WangZG-4430}, we study the $C\gamma_5\otimes \gamma_\mu C-C\gamma_\mu\otimes \gamma_5C$ type axialvector tetraquark states with the QCD sum rules in details, and observe that the $Z_c(3900)$ and $Z(4430)$ can be assigned to be the ground state and the first radial excited state of the axialvector tetraquark states with $J^{PC}=1^{+-}$, respectively \cite{Maiani-II-type,Z4430-Nielsen}, the energy gap between the $Z(4430)$ and the $Z_c(3900)$ is $576\,\rm{MeV}$.
For more works on this subject via QCD sum rules, one can consult Ref.\cite{Azizi-Zc}.
In Refs.\cite{Wang-3915-CgmCgm,Wang-3915-C5C5}, we study the $C\gamma_\mu\otimes \gamma^\mu C$-type and
$C\gamma_5\otimes \gamma_5 C$-type $cs\bar{c}\bar{s}$ scalar tetraquark states with the QCD sum rules in a systematic way,
and observe that  the $X(3915)$ and $X(4500)$ can be assigned to be the ground state and the first radial excited state of the scalar tetraquark states respectively, the energy gap between the $X(4500)$ and the $X(3915)$ is $588\,\rm{MeV}$. In this article, we will take the continuum threshold parameters as $\sqrt{s_0}=M_{Y}+(0.55\pm 0.10)\,\rm{GeV}$.

Now we search for the ideal  Borel parameters $T^2$ and continuum threshold parameters $s_0$  to satisfy   the  following four criteria:\\
$\bf 1.$ Pole dominance at the phenomenological side;\\
$\bf 2.$ Convergence of the operator product expansion;\\
$\bf 3.$ Appearance of the Borel platforms;\\
$\bf 4.$ Satisfying the modified  energy scale formula,\\
  via try and error, and  obtain the Borel parameters or Borel windows $T^2$, continuum threshold parameters $s_0$, ideal energy scales of the QCD spectral densities, pole contributions of the ground states, and contributions of the vacuum condensates of dimension  $10$, which are  shown   explicitly in Table 3.

From Table 3, we can see that the pole dominance at the phenomenological side is well satisfied, the operator product expansion is well convergent.
We take into account all uncertainties of the input parameters,
and obtain the values of the masses and pole residues of
 the   vector tetraquark states, which are  shown explicitly in Table 4 and in Figs.1-2.
From Figs.1-2, we can see that  there appear platforms in the Borel windows.  From Tables 3-4, we can see that the modified energy scale formula $\mu=\sqrt{M_{X/Y/X}^2-(4.1\,\rm{GeV})^2}$ can be well satisfied.  Now the four criteria of the QCD sum rules are all satisfied, and we expect to make reliable predictions. In Fig.1, we also plot the predicted masses of the tetraquark states $|0, 0; 0, 1; 1\rangle$ and $|1, 1; 0, 1; 1\rangle$
from the QCD sum rules without including the contributions of the vacuum condensates  $\langle\frac{\alpha_{s}GG}{\pi}\rangle$, $\langle\bar{q}q\rangle\langle\frac{\alpha_{s}GG}{\pi}\rangle$ and $\langle\bar{q}q\rangle^2\langle\frac{\alpha_{s}GG}{\pi}\rangle$, from the figure, we can see that
those contributions can be neglected approximately in the Borel windows. The predicted masses of the tetraquark states $\frac{1}{\sqrt{2}}\left(|1, 0; 1, 1; 1\rangle+|0, 1; 1, 1; 1\rangle\right)$ and $|1, 1; 2, 1; 1\rangle$  without including the contributions of the vacuum condensates  $\langle\frac{\alpha_{s}GG}{\pi}\rangle$, $\langle\bar{q}q\rangle\langle\frac{\alpha_{s}GG}{\pi}\rangle$ and $\langle\bar{q}q\rangle^2\langle\frac{\alpha_{s}GG}{\pi}\rangle$ are expected to be robust.

In Table 5, we present the possible assignments of the vector tetraquark states based on the QCD sum rules  compared to the assignments suggested in Ref.\cite{Ali-Maiani-Y}.

 The predicted mass $M_{Y}=4.24\pm0.10\,\rm{GeV}$ of the $|0, 0; 0, 1; 1\rangle$ tetraquark  state is in excellent agreement with the experimental value  $M_{Y(4220)}=4222.0\pm3.1\pm 1.4\,  \rm{MeV}$ from the BESIII    collaboration \cite{BES-Y4220-Y4320}, or the experimental value  $M_{Y(4260)}=4230.0\pm 8.0\,  \rm{MeV}$ from Particle Data Group \cite{PDG}, which supports  assigning the $Y(4260/4220)$  to be the  $C\gamma_5\otimes\stackrel{\leftrightarrow}{\partial}_\mu\otimes \gamma_5C$   type vector tetraquark state.

 The predicted mass $M_{Y}=4.28\pm0.10\,\rm{GeV}$ of the $|1, 1; 0, 1; 1\rangle$ tetraquark state is compatible  with the experimental values  $M_{Y(4220)}=4222.0\pm3.1\pm 1.4\,  \rm{MeV}$ and $M_{Y(4320)}=4320.0\pm 10.4 \pm 7.0\, \rm{MeV}$ from the BESIII    collaboration \cite{BES-Y4220-Y4320}, or the experimental values  $M_{Y(4260)}=4230.0\pm 8.0\,  \rm{MeV}$ and $M_{Y(4360)}=4368.0\pm 13\,  \rm{MeV}$ from Particle Data Group \cite{PDG}, which supports  assigning the $Y(4260/4220)$ or the $Y(4360/4320)$ to be the  $C\gamma_\alpha\otimes\stackrel{\leftrightarrow}{\partial}_\mu\otimes \gamma^{\alpha}C$   type vector tetraquark state.

 The predicted masses $M_{Y}=4.31\pm0.10\,\rm{GeV}$ of the $\frac{1}{\sqrt{2}}\left(|1, 0; 1, 1; 1\rangle+|0, 1; 1, 1; 1\rangle\right)$ tetraquark state  and  $M_{Y}=4.33\pm0.10\,\rm{GeV}$ of the $|1, 1; 2, 1; 1\rangle$ tetraquark state are compatible  with the experimental values  $M_{Y(4320)}=4320.0\pm 10.4 \pm 7.0\, \rm{MeV}$ and $M_{Y(4390)}=4391.6\pm6.3\pm1.0\,\rm{MeV}$ from the BESIII    collaboration \cite{BES-Y4390,BES-Y4220-Y4320}, or the experimental value  $M_{Y(4360)}=4368.0\pm 13\,  \rm{MeV}$ from Particle Data Group \cite{PDG}, which supports  assigning the $Y(4360/4320)$ or the $Y(4390)$ to be the
 $C\gamma_\mu \otimes\stackrel{\leftrightarrow}{\partial}_\alpha \otimes\gamma^\alpha C +C\gamma^\alpha \otimes\stackrel{\leftrightarrow}{\partial}_\alpha \otimes\gamma_\mu C$ type or the $
C\gamma_5 \otimes\stackrel{\leftrightarrow}{\partial}_\mu \otimes\gamma_\nu C
+C\gamma_\nu \otimes\stackrel{\leftrightarrow}{\partial}_\mu \otimes\gamma_5 C
-C\gamma_5 \otimes\stackrel{\leftrightarrow}{\partial}_\nu \otimes\gamma_\mu C
-C\gamma_\mu \otimes\stackrel{\leftrightarrow}{\partial}_\nu \otimes\gamma_5 C $ type  vector tetraquark states.

The present predictions disfavor assigning the   $Y(4660)$ to be the $C\gamma_5\otimes\stackrel{\leftrightarrow}{\partial}_\mu\otimes \gamma_5C$   type, $C\gamma_\alpha\otimes\stackrel{\leftrightarrow}{\partial}_\mu\otimes \gamma^{\alpha}C$   type, $C\gamma_\mu \otimes\stackrel{\leftrightarrow}{\partial}_\alpha \otimes\gamma^\alpha C +C\gamma^\alpha \otimes\stackrel{\leftrightarrow}{\partial}_\alpha \otimes\gamma_\mu C$ type or  $
C\gamma_5 \otimes\stackrel{\leftrightarrow}{\partial}_\mu \otimes\gamma_\nu C
+C\gamma_\nu \otimes\stackrel{\leftrightarrow}{\partial}_\mu \otimes\gamma_5 C
-C\gamma_5 \otimes\stackrel{\leftrightarrow}{\partial}_\nu \otimes\gamma_\mu C
-C\gamma_\mu \otimes\stackrel{\leftrightarrow}{\partial}_\nu \otimes\gamma_5 C $ type  vector tetraquark states. While in Ref.\cite{Ali-Maiani-Y}, the $Y(4660)$ is assigned to be tetraquark state $|1, 1; 2, 1; 1\rangle$ by fitting the experimental values of the masses with the diquark-antidiquark model. Our previous calculations based on the QCD sum rules indicate that the $Y(4660)$ can be assigned to be the $C \otimes \gamma_\mu C$ type vector tetraquark state $c\bar{c}s\bar{s}$ \cite{WangY4360Y4660-1803} or the $C\gamma_\mu \otimes \gamma_\nu C-C\gamma_\nu \otimes \gamma_\mu C$ type vector tetraquark state  $c\bar{c}q\bar{q}$ \cite{WangEPJC-1601}, where the relative P-waves  lie in the diquarks or antidiquarks.

\begin{table}
\begin{center}
\begin{tabular}{|c|c|c|c|c|c|c|c|}\hline\hline
$|S_{qc}, S_{\bar{q}\bar{c}}; S, L; J\rangle$                                &$\mu(\rm{GeV})$ &$T^2(\rm{GeV}^2)$ &$\sqrt{s_0}(\rm{GeV})$ &pole    &$D(10)$ \\ \hline

$|0, 0; 0, 1; 1\rangle$                                                      &$1.1$         &$2.2-2.8$       &$4.80\pm0.10$        &$(49-81)\%$   &$\leq 1\%$ \\ \hline

$|1, 1; 0, 1; 1\rangle$                                                      &$1.2$         &$2.2-2.8$       &$4.85\pm0.10$        &$(45-79)\%$   &$(1-5)\%$ \\ \hline

$\frac{1}{\sqrt{2}}\left(|1, 0; 1, 1; 1\rangle+|0, 1; 1, 1; 1\rangle\right)$ &$1.3$         &$2.6-3.2$       &$4.90\pm0.10$        &$(46-75)\%$   &$\ll 1\%$ \\ \hline

$|1, 1; 2, 1; 1\rangle$                                                      &$1.4$         &$2.6-3.2$       &$4.90\pm0.10$        &$(40-71)\%$   &$\leq 1\%$ \\ \hline \hline
\end{tabular}
\end{center}
\caption{ The Borel windows $T^2$, continuum threshold parameters $s_0$, ideal energy scales of the QCD spectral densities, pole contributions of the ground states, and contributions of the vacuum condensates of dimension  $10$.   }
\end{table}

\begin{table}
\begin{center}
\begin{tabular}{|c|c|c|c|c|c|c|c|}\hline\hline
$|S_{qc}, S_{\bar{q}\bar{c}}; S, L; J\rangle$                                &$M_Y(\rm{GeV})$  &$\lambda_Y(10^{-2}\rm{GeV}^6)$   \\ \hline

$|0, 0; 0, 1; 1\rangle$                                                      &$4.24\pm0.10$    &$2.31 \pm0.45$            \\ \hline

$|1, 1; 0, 1; 1\rangle$                                                      &$4.28\pm0.10$    &$4.93 \pm1.00$              \\ \hline

$\frac{1}{\sqrt{2}}\left(|1, 0; 1, 1; 1\rangle+|0, 1; 1, 1; 1\rangle\right)$ &$4.31\pm0.10$    &$2.99 \pm0.54$               \\ \hline

$|1, 1; 2, 1; 1\rangle$                                                      &$4.33\pm0.10$    &$7.35 \pm1.39$               \\ \hline \hline
\end{tabular}
\end{center}
\caption{ The masses and pole residues of the vector tetraquark states.   }
\end{table}

\begin{table}
\begin{center}
\begin{tabular}{|c|c|c|c|c|c|c|c|}\hline\hline
$|S_{qc}, S_{\bar{q}\bar{c}}; S, L; J\rangle$                                &$M_Y(\rm{GeV})$     & This Work       &\cite{Ali-Maiani-Y} \\ \hline
$|0, 0; 0, 1; 1\rangle$                                                      &$4.24\pm0.10$       &$Y(4220)$        &$Y(4220)$             \\ \hline

$\frac{1}{\sqrt{2}}\left(|1, 0; 1, 1; 1\rangle+|0, 1; 1, 1; 1\rangle\right)$ &$4.31\pm0.10$       &$Y(4320/4390)$   &$Y(4330)$              \\ \hline

$|1, 1; 0, 1; 1\rangle$                                                      &$4.28\pm0.10$       &$Y(4220/4320)$   &$Y(4390)$             \\ \hline
$|1, 1; 2, 1; 1\rangle$                                                      &$4.33\pm0.10$       &$Y(4320/4390)$   &$Y(4660)$              \\ \hline \hline
\end{tabular}
\end{center}
\caption{ The masses  of the vector tetraquark states and possible assignments.   }
\end{table}

\begin{figure}
 \centering
 \includegraphics[totalheight=5cm,width=7cm]{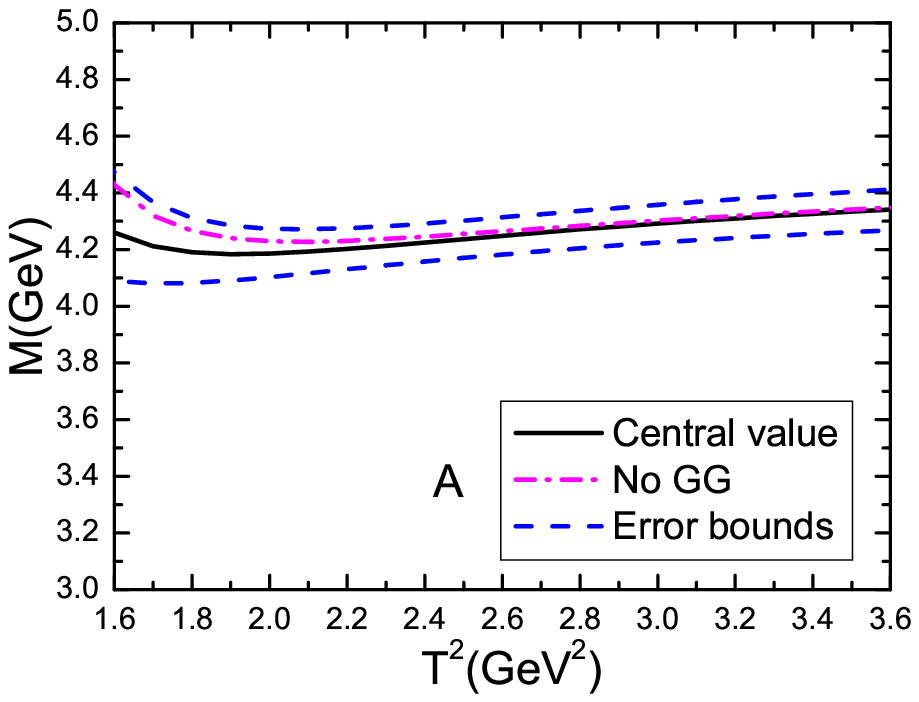}
 \includegraphics[totalheight=5cm,width=7cm]{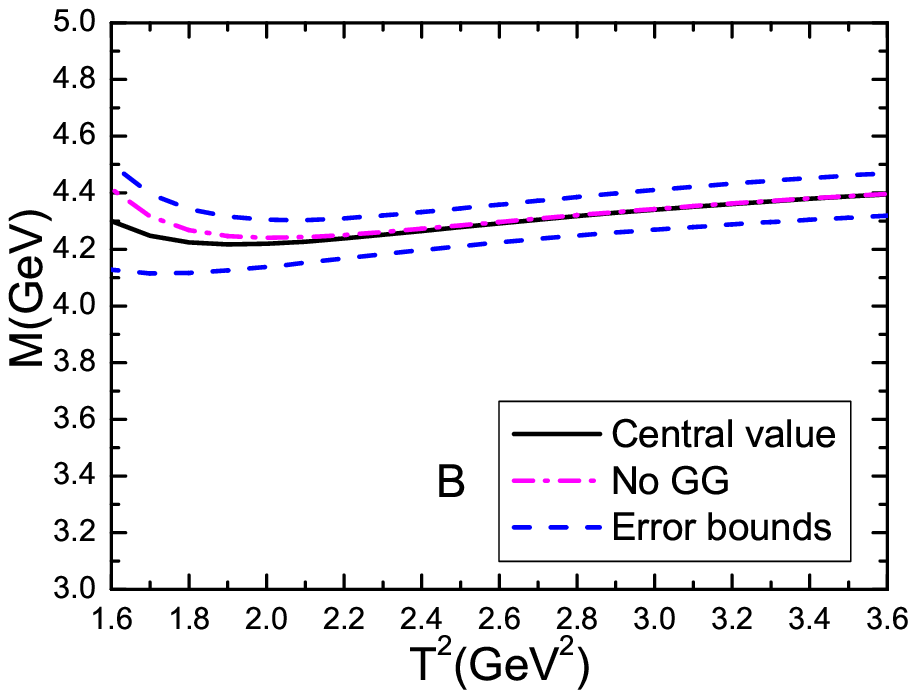}
 \includegraphics[totalheight=5cm,width=7cm]{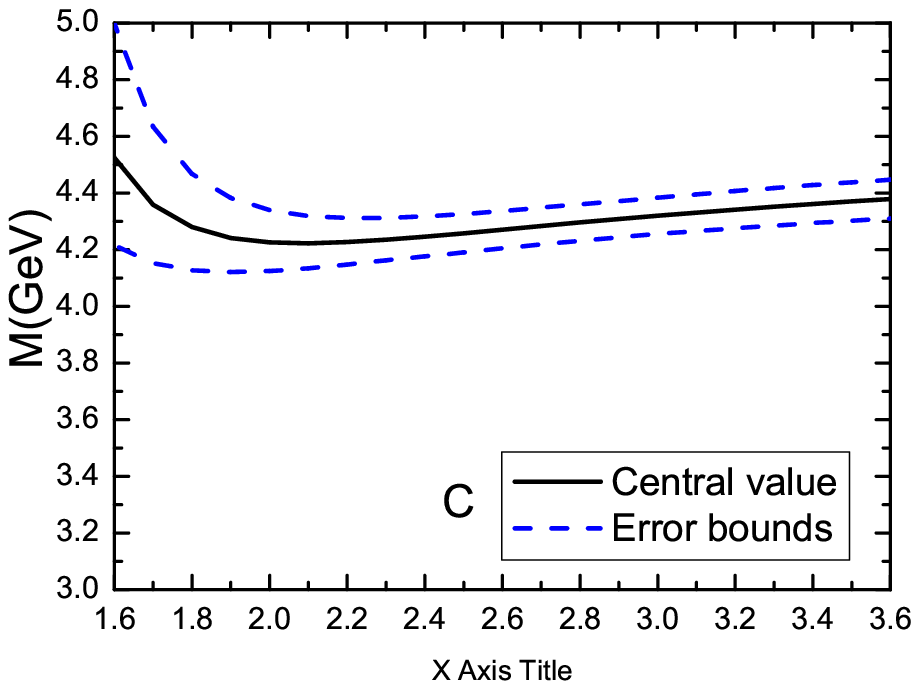}
 \includegraphics[totalheight=5cm,width=7cm]{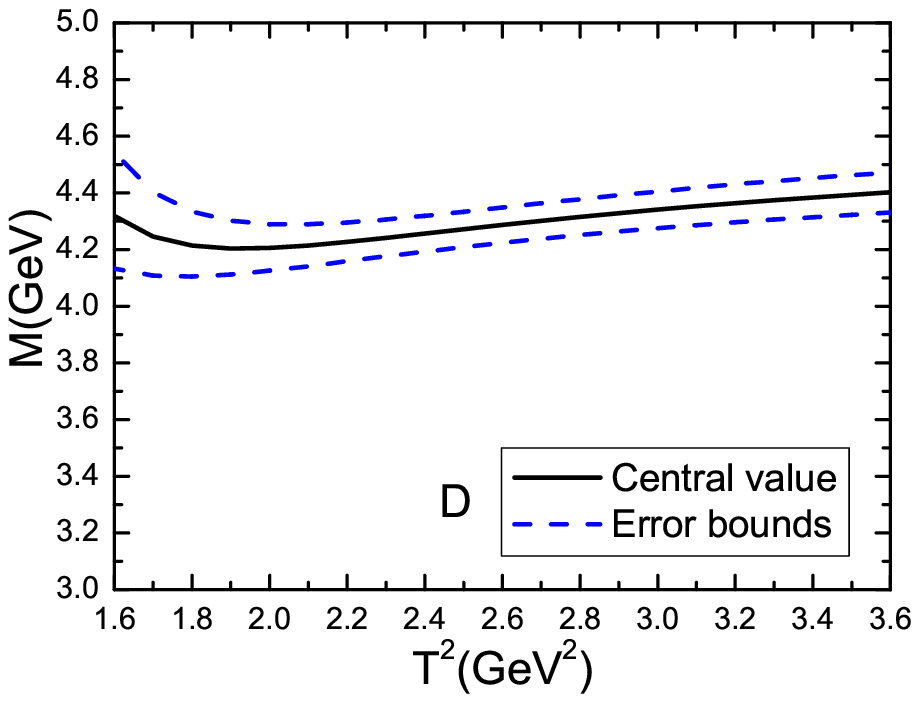}
  \caption{ The masses of the vector tetraquark states with variations of the Borel parameters $T^2$, where  the $A$, $B$, $C$ and $D$ denote the
  $|0, 0; 0, 1; 1\rangle$, $|1, 1; 0, 1; 1\rangle$, $\frac{1}{\sqrt{2}}\left(|1, 0; 1, 1; 1\rangle+|0, 1; 1, 1; 1\rangle\right)$ and $|1, 1; 2, 1; 1\rangle$ vector tetraquark states,  respectively, the "No GG" denotes the contributions of the vacuum condensates  $\langle\frac{\alpha_{s}GG}{\pi}\rangle$, $\langle\bar{q}q\rangle\langle\frac{\alpha_{s}GG}{\pi}\rangle$ and $\langle\bar{q}q\rangle^2\langle\frac{\alpha_{s}GG}{\pi}\rangle$ are excluded.   }
\end{figure}

\begin{figure}
 \centering
 \includegraphics[totalheight=5cm,width=7cm]{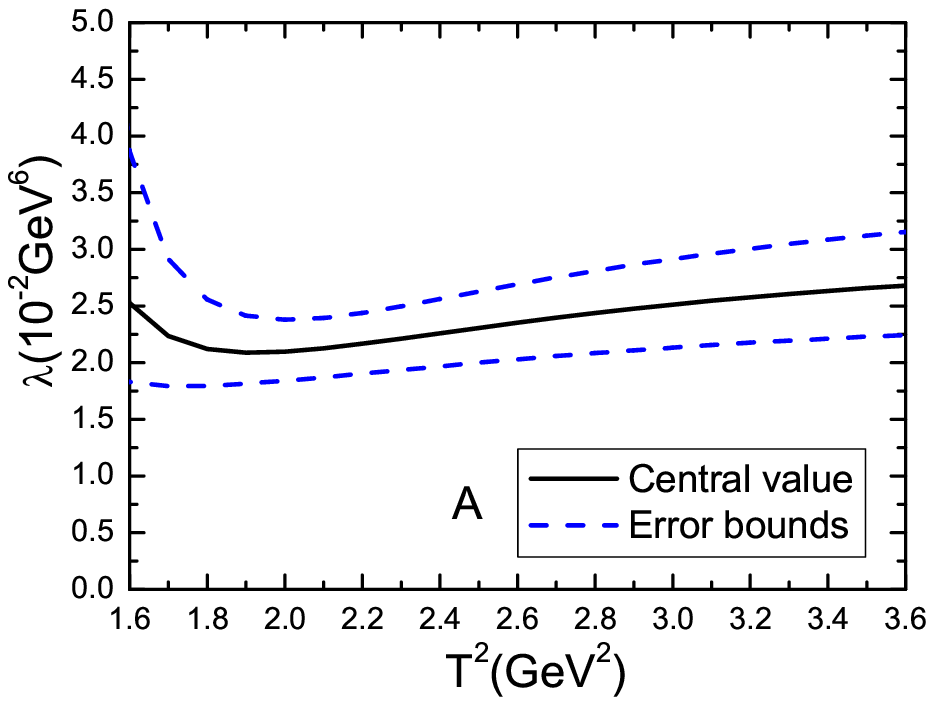}
 \includegraphics[totalheight=5cm,width=7cm]{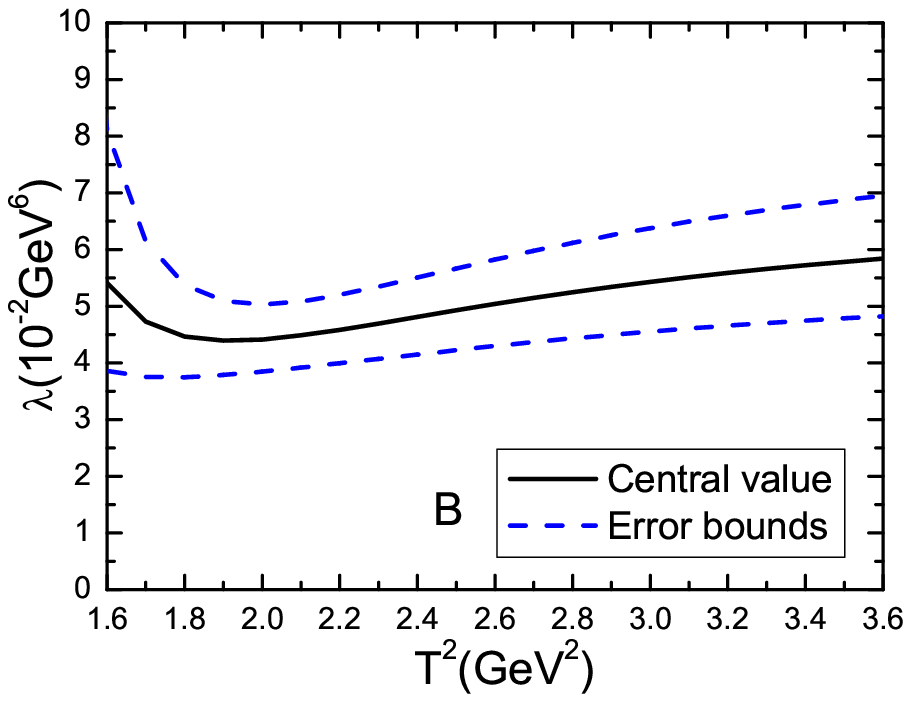}
 \includegraphics[totalheight=5cm,width=7cm]{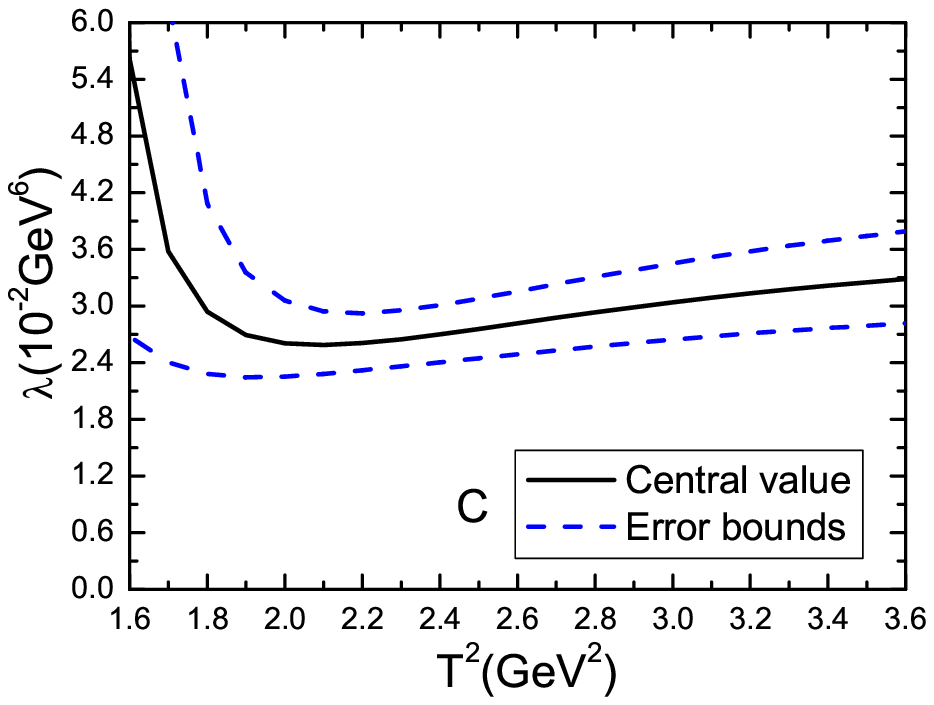}
 \includegraphics[totalheight=5cm,width=7cm]{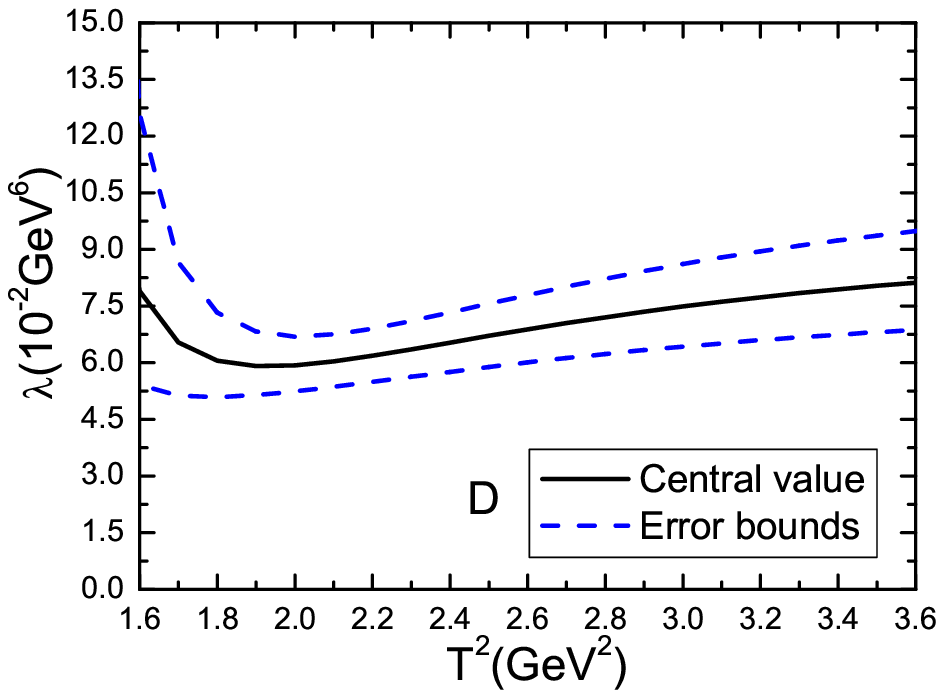}
  \caption{ The pole residues of the vector tetraquark states with variations of the Borel parameters $T^2$, where  the $A$, $B$, $C$ and $D$ denote the
  $|0, 0; 0, 1; 1\rangle$, $|1, 1; 0, 1; 1\rangle$, $\frac{1}{\sqrt{2}}\left(|1, 0; 1, 1; 1\rangle+|0, 1; 1, 1; 1\rangle\right)$ and $|1, 1; 2, 1; 1\rangle$ vector tetraquark states,  respectively.   }
\end{figure}

In 2008, the Belle Collaboration  observed two resonance-like structures ($Z(4050)$ and $Z(4250)$)  in the $\pi^+\chi_{c1}$ invariant mass
distribution  in the exclusive $\bar{B}^0\to K^- \pi^+ \chi_{c1}$ decays  with the statistical significances
exceeds $5 \sigma$, including the effects of systematics from various fit models \cite{Belle-2008-Z}.  The Breit-Wigner masses and widths are 
$M_{Z(4050)}=4051\pm14^{+20}_{-41} \,\rm{MeV}$, $\Gamma_{Z(4050)}=82^{+21}_{-17}$$^{+47}_{-22}\, \rm{MeV}$,
$M_{Z(4250)}=4248^{+44}_{-29}$$^{+180}_{-35}\,\rm{MeV}$ and
$\Gamma_Z(4250)=177^{+54}_{-39}$$^{+316}_{-61}\,\rm{MeV}$, respectively. If the $Z(4050)$ and $Z(4250)$ are really resonances, their
 quark contents must be  $c\bar{c} u\bar{d}$ according to the non-zero electronic charge. If the $Z(4050)$ and $Z(4250)$ are scalar tetraquark states, the decays $Z(4050/4250) \to \pi^+\chi_{c1}$ take place through the relative P-wave; on the other hand, if they are vector tetraquark states, the decays
take place  through the relative S-wave.  The predicted masses $4.24\pm0.10\,\rm{GeV}$, $4.31\pm0.10\,\rm{GeV}$, $4.28\pm0.10\,\rm{GeV}$ and $4.33\pm0.10\,\rm{GeV}$
 for the vector tetraquark states $|0, 0; 0, 1; 1\rangle$, $\frac{1}{\sqrt{2}}\left(|1, 0; 1, 1; 1\rangle+|0, 1; 1, 1; 1\rangle\right)$,
$|1, 1; 0, 1; 1\rangle$  and
$|1, 1; 2, 1; 1\rangle$  respectively are all consistent with the  experimental data $M_{Z(4250)}=4248^{+44}_{-29}$$^{+180}_{-35}\,\rm{MeV}$  from the Belle Collaboration
considering the large uncertainties. The present predictions support assigning the $Z(4250)$ to the vector tetraquark state with a relative P-wave between  the diquark and antidiquark pair.

We cannot identify  a particle unambiguously with the mass alone, we have to study the decays of the $Y(4260/4220)$, $Y(4360/4320)$, $Y(4390)$ and $Y(4660/4630)$ with the QCD sum rules to testify the assignments in the scenario   of the  tetraquark states, it is our next work.
Experimentally, a number of  decays of the $Y(4260/4220)$, $Y(4360/4320)$, $Y(4390)$ and $Y(4660/4630)$ have been observed, such as
\begin{eqnarray}
Y(4220)&\to& \omega\chi_{c0}\, , \, J/\psi\pi^+\pi^-\, , \,  h_c\pi^+\pi^-  \, ,\nonumber \\
Y(4260)&\to& X(3872)\,\gamma \, , \, Z_c(3900)^+\pi^-\, ,\nonumber \\
Y(4320)&\to& J/\psi\pi^+\pi^-\, , \,\psi^\prime \pi^+\pi^-   \, ,\nonumber \\
Y(4390)&\to& h_c\pi^+\pi^-   \, ,\nonumber \\
Y(4660)&\to& \psi^\prime \pi^+\pi^- \, ,\, \Lambda_c^+ \Lambda_c^-    \, .
\end{eqnarray}
For detailed reviews on the properties of the $X$, $Y$, $Z$ states, one can consult the Refs.\cite{Esposito-PRT,Review-XYZ}.

\section{Conclusion}
In this article, we introduce the relative P-wave between the diquark and antidiquark explicitly to construct the vector tetraquark currents,
 then carry out the operator product expansion up to the vacuum condensates of dimension 10, take the modified energy scale formula to determine the optimal energy scales of the QCD spectral densities, and study the masses and pole residues of the vector tetraquark states  with the QCD sum rules systematically. We obtain the lowest vector tetraquark masses up to now, the present predictions support assigning the
 $Y(4220/4260)$,  $Y(4320/4360)$, $Y(4390)$ and $Z(4250)$ to be the vector tetraquark   states with a relative P-wave between the diquark and antidiquark pair.

\section*{Acknowledgements}
This  work is supported by National Natural Science Foundation, Grant Number  11775079.

\end{document}